# Material and radiation losses in structures with bound states in the continuum: $Ge_2Sb_2Te_5$ as an example

*Daria V. Bochek, Nikolay S. Solodovchenko, Denis A. Yavsin, Alexander B. Pevtsov, Kirill B. Samusev, and Mikhail F. Limonov\**

D. V. Bochek, N. S. Solodovchenko, Dr. K. B. Samusev, Prof. M. F. Limonov
Department of Physics and Engineering, ITMO University, St. Petersburg, 197101, Russia
m.limonov@metalab.ifmo.ru

Dr. D. A. Yavsin, Dr. A. B. Pevtsov, Dr. K. B. Samusev, Prof. M. F. Limonov
Ioffe Institute, St. Petersburg 194021, Russia.

The problem of losses in resonators is a key issue in the design and use of modern plasmonic and photonic meta-devices. We propose a spectroscopic method for determining material and radiation losses in structures with bound states in the continuum using as example the phase-change chalcogenide $Ge_2Sb_2Te_5$. The complex refractive index for the amorphous and crystalline $Ge_2Sb_2Te_5$ phases was measured experimentally, and the resonance optical properties were calculated for a telecommunication wavelength of 1500 nm. To demonstrate the validity of the method, two types of resonators were considered - a cylinder and a ring which support bound states in the continuum according to the Friedrich - Wintgen mechanism. We have determined the value of the material quality-factor for the amorphous $Ge_2Sb_2Te_5$ as $Q_{mat} = 46$ and the values of the radiation quality-factors $Q_{rad}$=608 for the cylinder and $Q_{rad}$=332 for the ring of a pair of $TE_{0,1,2}$ and $TE_{0,2,0}$ modes in the regime of bound states in the continuum. We also demonstrated two mechanisms for efficient switching of the scattering spectra between the amorphous and crystalline phases of $Ge_2Sb_2Te_5$.

# 1. Introduction

Future technologies are targeting a dramatic increase in subwavelength photonic integration, far exceeding those of bulk optical components and silicon photonics. An important step along this path was the concepts of metamaterials, metasurfaces, and nanophotonics, based on the structuring of materials at subwavelength scale.[1,2] In the implementation of the concept of metamaterials, most of the created structures with a magnetic response contained metals that have high Ohmic losses at optical frequencies, which limit the efficiency and any useful characteristics. Recently, it was experimentally demonstrated that subwavelength dielectric nanoparticles have induced magnetic multipoles due to Mie resonances and, unlike plasmonic metal particles, they do not suffer from Ohmic losses due to the absence of free charges.[3-5] Unfortunately, this advantage of dielectric particles does not solve all material loss problems, since any linear and causal material necessarily contains losses, as follows from the Kramers-Kronig relations.[6] In addition to material losses, the efficiency of the resonator are reduced by radiation losses and losses associated with defects and imperfect shape of the resonant particle. If material losses are a characteristic of a defect-free crystal structure, then radiation losses can be reduced by various methods, including the implementation of the Anderson localization regime,[7] the use of photonic crystals with a complete photonic band gap,[8] and the regime of bound states in the continuum.[9]

Bound states in the continuum (BIC) is an exclusive light trapping and confinement mechanism, a wave phenomenon that was first mathematically proposed in 1929 by von Neumann and Wigner[10] for electronic states. Although the von Neumann and Wigner model was never implemented in practice, the idea itself turned out to be very fruitful, as a result, other mechanisms of BIC formation were theoretically proposed and experimentally demonstrated[11-19] for various types of waves, including electromagnetic, acoustic, elastic and water waves. Photonic BICs coexist with propagating electromagnetic waves in a continuum, but theoretically remain completely localized in the structure without any radiation and are characterized by an infinite Q-factor.[9] In real structures, due to finite dimensions, material losses, and a number of other reasons, photonic bound states have a finite lifetime, finite Q-factor and are called quasi-BIC.[18] There are a number of physical mechanisms that lead to the formation of BIC;[9] in this work, we will be interested in quasi-BICs, which are achieved by adjusting the structural parameters of the material. This type of quasi-BIC arises when two non-orthogonal photonic modes are coupled to the same radiation channel, and a regime of avoided crossing arises at the appropriate structural parameters. This regime is described by the Friedrich–Wintgen model[20] when due to destructive interference in the far field zone one of the emitting modes disappears and becomes a quasi-BIC. Recently, the Friedrich-Wintgen BIC was discovered and studied in detail in dielectric cylinders and rings, when two eigenmodes with different polarizations associated with Mie- and Fabry-Perot- resonances strongly interact (strongly[21,22] or weakly[23]) near the avoided crossing regime.

The aim of this work is to separate material and radiation losses in resonators supporting quasi-BIC and formed from realistic material that is used in modern devices. For the calculations, we chose resonators in the form of a cylinder and a ring, and a phase-change material was used as a model material. Among phase-change materials, a special place is occupied by chalcogenide Ge-Sb-Te alloys, including the most commercialized compound $Ge_2Sb_2Te_5$ (hereinafter GST).[24-30] Thus, we get the opportunity not only to study the quasi-BIC in GST resonators, but also to trace their transformation during the phase transition. This material can quickly and drastically switch between a disordered amorphous state and an ordered crystalline state, usually via an electrical or optical pulse. In this case, in the infrared spectral range, such switching occurs between two dielectric phases, and in the optical range closer to ultraviolet, a transition occurs between the dielectric amorphous phase and the crystalline metal phase. Reversible switching between phase states is non-volatile and can be achieved on the nano-, pico-, and even femtosecond timescale.[31,32] Owing to these flexibilities, GST can be used as an active material in storage devices and to tune many properties, applications include control of radiation,[24] energy transfer,[25] resonance characteristics of a nanoantenna,[26] wavefront switching[27] among other applications.[28-30]

The importance of reducing material losses in structures with quasi-BIC was noted in a number of works.[33-41] In particular, for a number of materials this problem was discussed in ref. [33]. All-dielectric photonic crystal structures that are able to sustain effective near-zero refractive index modes coupled to quasi-BIC have been investigated theoretically[34] and experimentally.[35] A subwavelength core-shell particle made of a material with infinite or zero dielectric constant that supports quasi-BIC was studied.[36] Unfortunately, materials with such extreme refractive indices for optical frequencies have not yet been realized and researchers have to deal with conventional materials with finite losses. Therefore, the question of the numerical determination of material losses and their relationship with the value of radiation losses of a particular resonator remained open as before. This article is devoted to solving this problem based on the analysis of spectroscopic data.

## 2. Experiment: Complex Refractive Index

To numerically investigate the photonic properties of GST and, in particular quasi-BIC, it is necessary to know the complex permittivity $\varepsilon=\varepsilon'+i\varepsilon''$ (where $\varepsilon'=\text{Re}\varepsilon$, $\varepsilon''=\text{Im}\varepsilon$) for the crystalline and amorphous phases. These parameters can be determined on the basis of the experimentally measured complex refractive index $N=n+ik$, where $n$ − refractive index, $k$ − extinction coefficient. These measurements were carried out on two GST films prepared by laser electrodispersion technique,[42] which were amorphous GST films with a thickness of ~ 50 nm on a quartz substrate, ground from the back side. To obtain the GST film in the crystalline state, one of the samples was annealed at 170°C for 1 hour. The ellipsometric spectra were collected at room temperature at different angles of incidence using an ellipsometer M-2000 J.A.Woollam. The processing of the obtained ellipsometric data was carried out using the supplied CompleteEASE software. To determine the complex refractive index (**Figure 1**a,b), the standard model of an absorbing film on a semi-infinite quartz substrate was used, and the surface roughness of the GST film was also taken into account. It can be seen that the crystalline phase has higher values of n and k than the amorphous phase, which corresponds to the differences in the chemical bond in GST before and after the phase transition.[43,44] Note that the spectral dependences of the optical constants n and k for the GST films obtained under our technological conditions are in reasonable agreement with the literature data.[45] Based on these data, the spectral dependences of the complex dielectric permittivity of GST in the crystalline and amorphous phases were calculated as $\varepsilon = N^2$, thus $\text{Re}\varepsilon = n^2 - k^2$ and $\text{Im}\varepsilon = 2nk$ (Figure 1c,d). Here, we will analyze the photonic properties of GST at the telecommunication wavelength 1.5 μm, at which the dielectric constant of the crystalline phase is 41.6 + i17.5, and that of the amorphous phase is 19.0+i0.42.

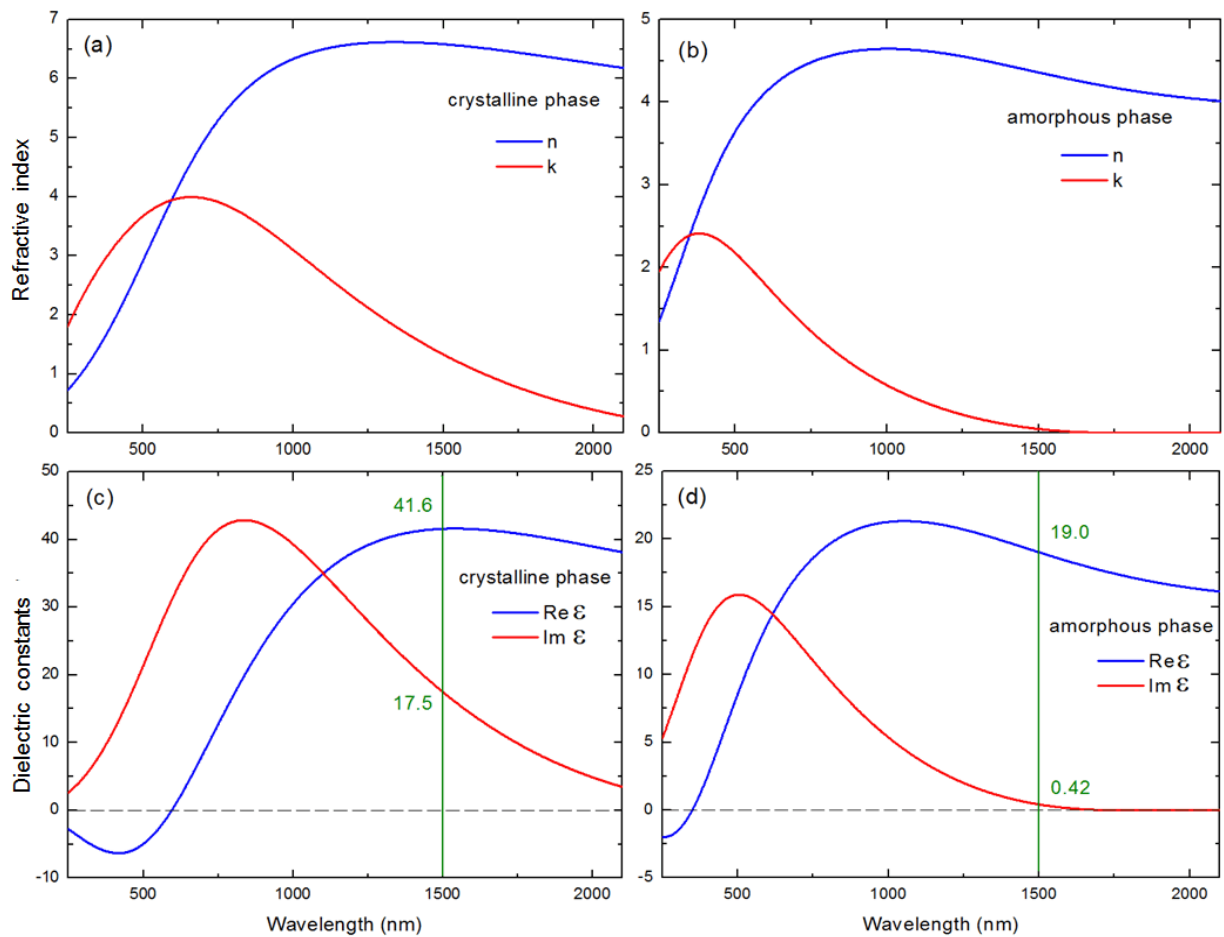

**Figure 1.** Experimentally measured complex refractive index $N = n + ik$, $n$ – refractive index, $k$ – extinction coefficient) of the GST film in crystalline (a) and amorphous (b) phases. Calculated complex permittivity ε=ε'+iε'' of the GST film in crystalline (c) and amorphous (d) phases.

## 3. Calculation Method

We present the results of a study of photonic peculiarities in the region of quasi-BIC in GST taking into account material losses. To study the transformation of quasi-BIC when the imaginary part of the permittivity changes from 0 (in the case of no material losses) to the experimentally obtained value, we used numerical calculations, which provide key information about the optical spectrum with resonant frequencies of eigenmodes and mode Q-factors. For the amorphous phase, we performed systematic calculations for a large set of ideal dielectric cylindrical and ring resonators with a uniform dielectric permittivity ε=ε'+iε'', where ε' = 19.0, and ε'' varies from 0 to 0.42 with a step of 0.02. For the crystalline phase, we consider resonators with a uniform dielectric constant 41.6 + i17.5. The environment was vacuum $\varepsilon_{vac} = 1$.

We have performed systematic calculations of the scattering cross-section (SCS) of cylinders with an outer radius r and a length l and rings with an inner radius $r_{in}$, an outer radius r and a length l. In the rings, the ratio of the radii was constant and amounted to $r_{in}/r = 0.4$. The *r/l* aspect ratio was varied in such a range to observe a certain quasi-BIC formed by the anticrossing of the two resonant modes $TE_{0,1,2}$ and $TE_{0,2,0}$. The results of the study of such a quasi-BIC in a cylinder with ε=80+i0 are presented in Ref.21, where the standard nomenclature[46] of modes of a cylindrical resonator $TE_{m,s,p}$ and $TM_{m,s,p}$ is used. In these notations, *m, s, p* are the indices denoting the azimuthal, radial, and axial indices, respectively. Only the modes with the same azimuthal index *m* could interact and we consider the case $m = 0$. We consider a scenario in which the structures are illuminated by a linearly polarized plane wave, the electric field is linearly polarized in the plane perpendicular to the axes of the cylinder and ring (TE polarization). For generality, when demonstrating the results we use the normalized size parameter $x = kr = r\omega/c = 2\pi r/\lambda$ being a product of the wavenumber $k$ and outer radius $r$.

All the computations of SCS were performed in the frequency domain using the commercial software COMSOL. In order to obtain sufficiently accurate solutions by numerical methods a physics-controlled mesh with the "extremely fine" option was used to capture the geometric details and to resolve the curvature of resonators boundaries.

## 4. Scattering Cross Section for Cylindrical and Ring Resonators of $Ge_2Sb_2Te_5$ in Amorphous Phase

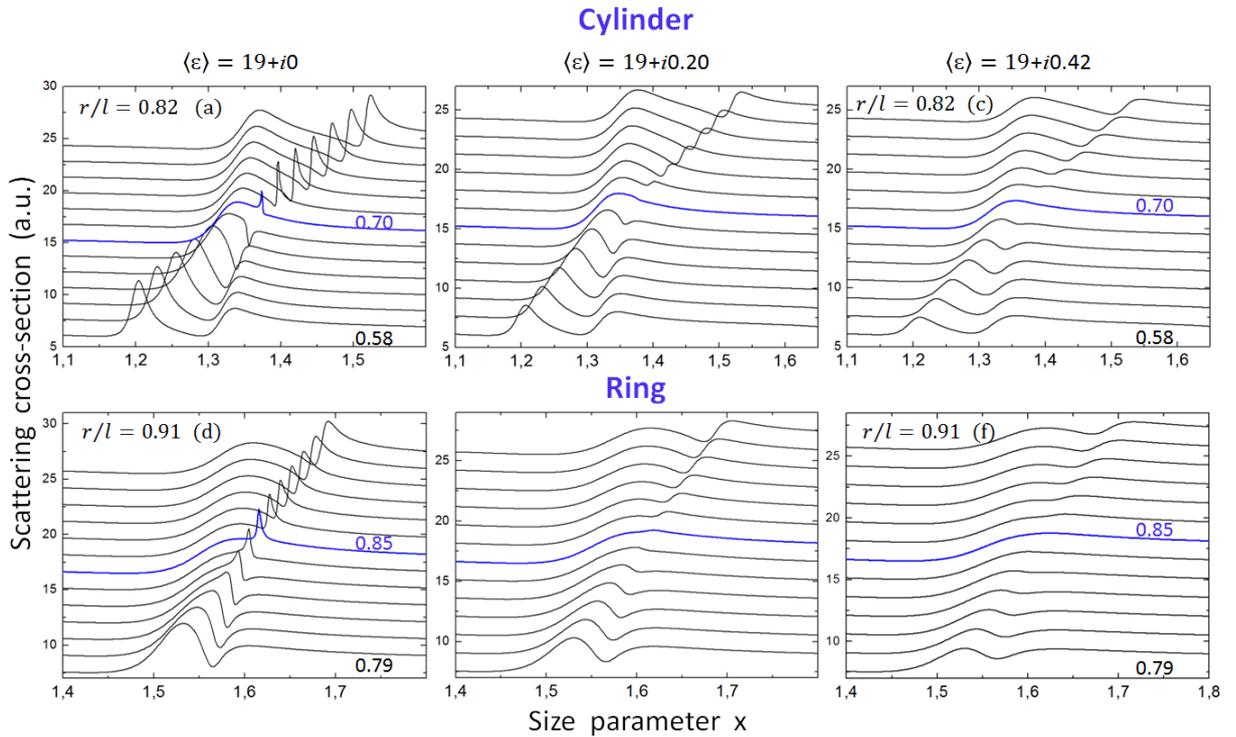

**Figure 2**. Spectra of the normalized SCS (for harmonic $m=0$) for cylindrical (a-c) and ring (d-f) of GST in amorphous phase as a function of their aspect ratio $r/l$ in the regions of avoided crossing regime between the Mie-type $TE_{0,2,0}$ and Fabry-Perot-type $TE_{0,1,2}$ modes. The spectra

marked in blue correspond to quasi-BIC. Normalized size parameter $x = kr = r\omega/c = 2\pi r/\lambda$. TE-polarized incident wave. The dielectric permittivity of each of the structures is indicated above the corresponding panel. The structures are placed in the vacuum, $\varepsilon_2 = 1$. Curves are shifted vertically by 2 a.u.

As demonstrated earlier in the cylinder as well as in the ring,[21-23] there are two types of modes with different behavior depending on the aspect ratio $r/l$. The modes of the first type are formed mainly due to reflection from the side wall, and they are associated with the Mie resonances of an infinite cylinder. Accordingly, Mie-type modes exhibit a small frequency shift with a change in the length of the structure. The modes of the second type are formed mainly due to reflection from two parallel faces of the cylinder or ring, and they could be associated with the Fabry-Perot modes. The Fabry-Perot-type modes demonstrate a strong shift to higher frequencies with increasing aspect ratio. Due to the different spectral shift of Mie and Fabry-Perot modes depending on the length of the structure $l$, for a fixed outer radius $r$, they should intersect at certain points in the parameter space ($r/l$, $x$). The modes with the same azimuthal index *m* interact and demonstrate strong coupling the avoided crossing regime at special values of the *r/l* parameter. Coupling creates a point of avoided crossing behavior, with the high frequency line nearly disappearing in the scattering spectra. The dramatic decrease in line intensity is the result of interference outside the resonator in accord with the Friedrich-Wintgen theory of BIC states. In addition, a complex interference between narrow lines and a wide background should be noted, which leads to Fano resonances[47] with asymmetric profiles of all quasi-Mie and quasi-Fabry-Perot lines.

**Figure 2** shows the SCS spectra of cylinders and rings in the avoided crossing regions of the $TE_{0,2,0}$ and $TE_{0,1,2}$ modes at three values of the dielectric permittivity 19.0+i0, 19.0+i0.20 and 19.0+i0.42. The latter value corresponds to amorphous GST. The spectra for the case of zero material losses (Figure 2a,d) clearly demonstrate that the distance of closest approach of the two lines occurs in the region $r/l = 0.70$ for a cylinder and 0.85 for a ring. This is the region of origin of quasi-BIC and it practically does not change with the addition of material losses for both the cylinder and the ring. It is clearly seen that with an increase in material losses, the intensity of both lines significantly decreases, while the half-width of the lines does not undergo dramatic changes.

## 5. Effect of Material Loss

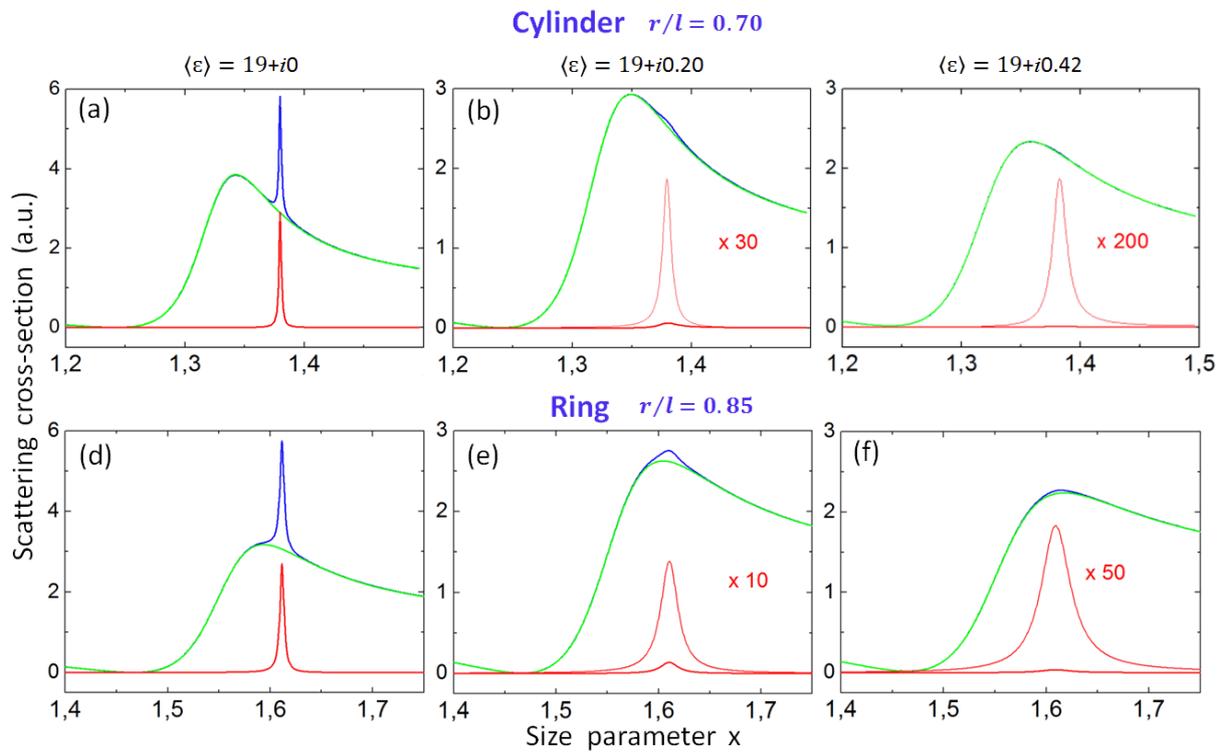

**Figure 3.** SCS spectra of the cylinder and ring for the aspect ratios r/l corresponding to quasi-BIC (blue line) (Figure 2), and their decomposition into two contours corresponding to the low-frequency (green line) and high-frequency (red line) branches. Normalized size parameter $x = r\omega/c$. TE-polarized incident wave.

Earlier, in a series of works, the Fano resonance in the scattering spectra of individual dielectric particles of various shapes was theoretically analyzed and experimentally demonstrated.[47,48] In particular, this concerns cylinders and rings, in the spectra of which a series of Fano resonances were observed, caused by the interference of two waves - the emitted resonant mode and nonresonant scattering on the whole object.[49,50] Fano resonance is observed as a result of the interference of two states, one of which is spectrally narrow and the other is wide when both states are excited by some external source. Note that there is a direct relationship between the Friedrich-Wintgen quasi-BIC and Fano resonances, since these two phenomena are associated with the same physical effect of wave interference. To analyze precisely quasi-BIC, we examine in detail the calculated SCS spectra from Figure 2 by decomposing them into two Fano profiles:

$$\text{SCS}(\omega) = A_1 \frac{(q_1 + \Omega_1)^2}{1 + \Omega_1^2} + A_2 \frac{(q_2 + \Omega_2)^2}{1 + \Omega_2^2}, \tag{1}$$

where $q$ is the Fano parameter, $\Omega = 2(\omega - \omega_0)/\Gamma$, where $\Gamma$ and $\omega_0$ are the resonance width and frequency respectively. The Fano profile is generally asymmetric and determined by the parameter q, which is the only new feature in the Fano profile in comparison with the Lorentzian profile. The position of the quasi - BIC is determined by the maximum $Q$-factor of the line and corresponds to the Fano parameter $q \to \infty$.[21,22] At this value, the Fano lineshape becomes a symmetric Lorentzian function and the resonance does not couple to the continuum of states.

The results of the decomposition of the spectra are shown in Figure 3 for a cylinder and a ring at three values of the dielectric permittivity. If one look at the SCS spectra corresponding to the material loss in amorphous GST (Figure 3c,f), then it seems that the high-frequency mode has become completely invisible. Although it does indeed resemble a dark state, the accurate decomposition of the SCS spectrum into two Fano contours allows one to observe a clear symmetric Lorentzian line corresponding to the quasi - BIC. From Figure 3, we obtain an important result: with an increase in material losses from 0 to 0.42, the low-frequency peak almost does not change in amplitude, while the high-frequency peak corresponding to the

quasi - BIC decreases in amplitude by approximately two orders of magnitude. Moreover, the decrease in the quasi - BIC amplitude in the spectrum of the cylinder is stronger than in the spectrum of the ring.

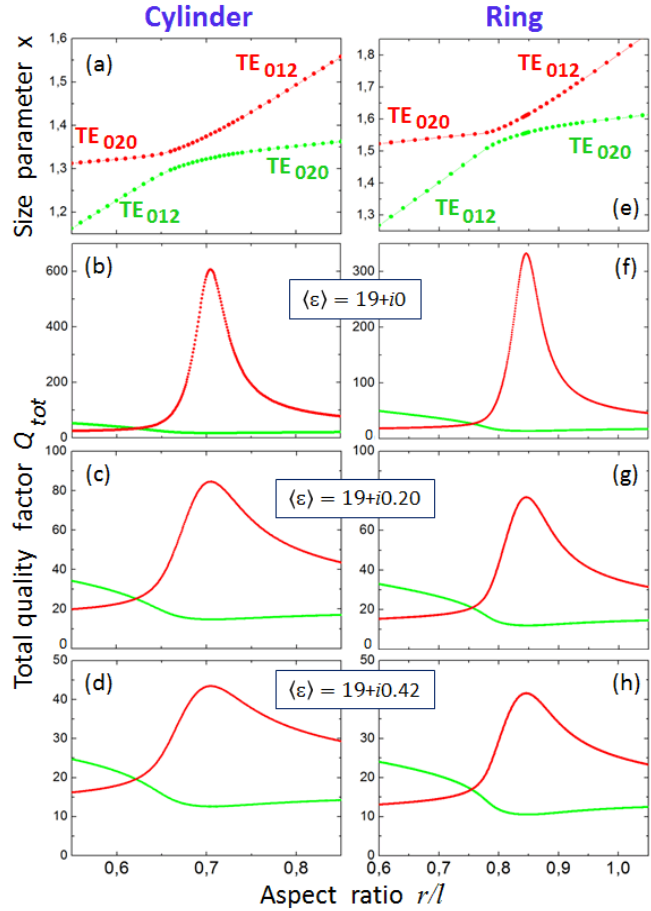

**Figure 4**. a, e) Calculated frequencies of the $TE_{0,1,2}$ and $TE_{0,2,0}$ modes in the avoided crossing regions for the cylinder and the ring as a function of its aspect ratio. Dependences of the total quality factor $Q_{tot}$ of the high-frequency branch (red) and low-frequency branch (green) on the aspect ratio for the cylinder (b-d) and ring (f-h) for three values of the dielectric permittivity. TE-polarized incident wave, normalized size parameter $x = r\omega/c$.

**Figure 4** demonstrates further results of processing the SCS spectra. Outside the avoided crossing regime the frequency shifts of both Mie and Fabry-Perot modes are well described by a linear law. As the frequencies approach each other, a classical resonance picture of the formation of quasi-BIC according to the Friedrich - Wintgen mechanism is observed. It should be noted that the position of both peaks practically did not change and the value of the minimum splitting (Rabi splitting) did not change when the imaginary part of the permittivity ε″ changed from 0 to 0.42. Therefore, in Figure 4, we present the frequency dependences only for ε″ = 0.

In the avoided crossing region (Figure 4a,e), the total quality factor $Q_{tot}$ of the high-frequency branch increases sharply, reaching in the quasi-BIC regime a value of 608 for a cylinder and 332 for a ring in the absence of losses in both resonators. As a result of the interaction of two modes, the low-frequency branch demonstrates the opposite behavior, namely, its $Q_{tot}$-factor has a minimum in the aspect ratio region corresponding to the quasi-BIC. Note that the quasi-BIC regions of the cylinder and the ring are observed at different values of the aspect ratio.

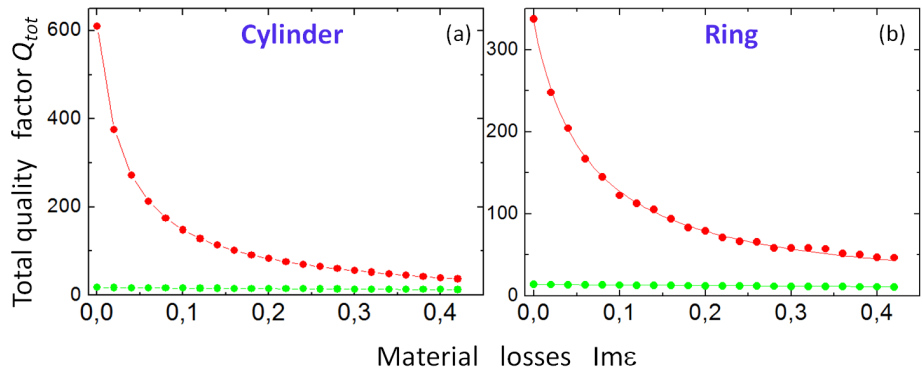

**Figure 5.** Dependences of the total quality factor $Q_{tot}$ of the high-frequency branch (red) and low-frequency branch (green) on the material losses for the cylinder (a) and ring (b).

The obtained results enable us to analyze the influence of material losses on the total quality factor $Q_{tot}$ of the quasi-BIC in the strong coupling regime. Figure 5 shows the dependence of the factor $Q_{tot}$ of the high-frequency and low-frequency modes at the frequency corresponding to the quasi-BIC, depending on material losses, the level of which is determined by the value of the imaginary part of the dielectric permittivity. It can be seen that the Q-factor of the high-frequency mode strongly depends on material losses, while the Q-factor of the low-frequency mode in the scales shown in the Figure 5 does not practically change. The total $Q$-factor is determined by both radiation and material losses and is expressed as:[22]

$$Q_{tot}^{-1} = Q_{rad}^{-1} + Q_{mat}^{-1} \qquad (2)$$

Here $Q_{rad}$ and $Q_{mat}$ are responsible for the radiation and material losses, respectively. In the limit of vanishing material losses, we have $Q_{tot}^{-1} = Q_{rad}^{-1}$ (Figure 4a,e), so we can determine the $Q_{mat}$ for amorphous GSTs, assuming $Q_{rad}$ to be independent of the losses in each of the resonators. Moreover, the material losses $Q_{mat}$ not depend on the shape of the resonator; therefore, determining the value of $Q_{mat}$ for a cylinder and a ring and comparing them is a good test for the validity of our calculations. We use the values of quality factors obtained from numerical calculations for both resonators, namely for the cylinder $Q_{rad}=608$ and $Q_{tot}^{\varepsilon''=0.42} = 43$ and for the ring $Q_{rad}=332$ and $Q_{tot}^{\varepsilon''=0.42} = 41$. Finally, we get an impressive result that in both cases for an amorphous GST $Q_{mat} = 46$.

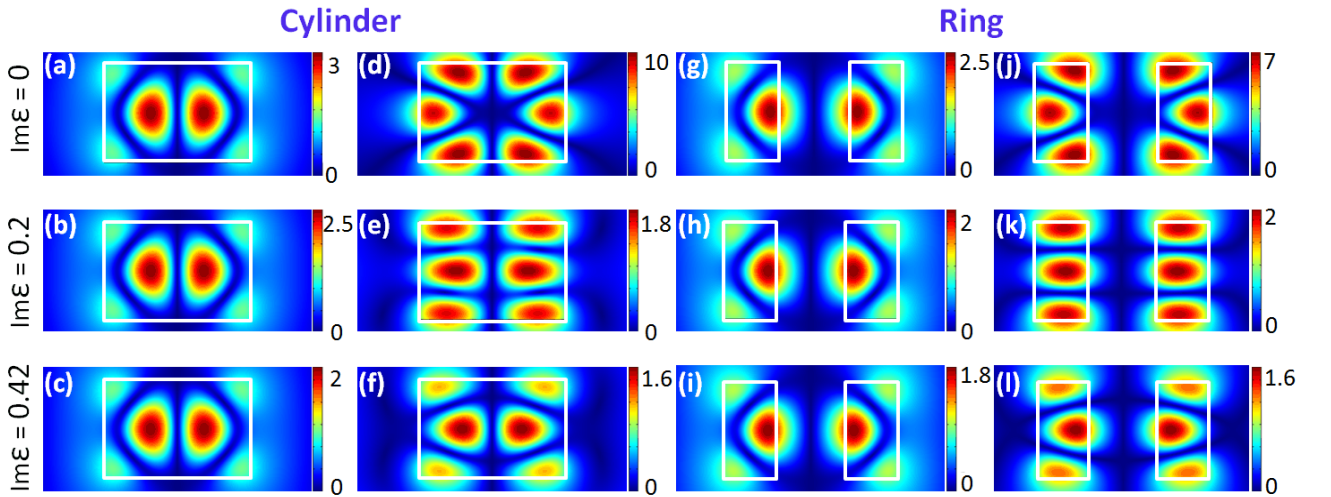

**Figure 6.** Distribution of the electric field amplitude $|E|$ in the (x, z) cross-sections (resonator side view) for the cylinder (a-f) and ring (g-l) for the low-frequency (a-c and g-i) and high-

frequency (d-f and j-l) branches at quasi-BIC aspect ratios for three values of losses shown in the picture on the left.

To provide a complete picture of the quasi-BIC transformation with a change in material losses in amorphous GST, we performed calculations of the distribution of the electric field amplitude |E| in the (x, z) cross-sections of the cylinder and ring. From **Figure 6** it is clearly seen that the distribution of the electric field amplitude in the cylinder and ring has a similar form with the difference that in the case of a ring, the field predominantly remains in the material with a high dielectric constant, shifting from the central cylindrical air region. Note that the field distribution of the low-frequency mode at a frequency corresponding to the quasi-BIC practically does not change its shape when the losses change from 0 to 0.42, but only slightly decreases in amplitude both in the cylinder and in the ring. This behavior fully correlates with changes in the intensity of the spectra shown in Figure 3. In contrast to the low-frequency mode, the high-frequency mode corresponding to the quasi-BIC changes significantly both in the amplitude and in the field distribution. In the absence of losses, the field distribution contains six almost identical lobes and, in the case of a cylinder, is well described by the $C_6$ symmetry, Figure 6d. This distribution is the result of the interaction of the Mie-type $TE_{0,2,0}$ mode with a transverse field distribution and the Fabry-Perot $TE_{0,1,2}$ mode with a longitudinal (along the vertical axis) distribution of the field. It is important to note that the intensities of the lines of both modes in the SCS spectra are approximately equal, Figure 3a,d. This distribution was previously presented in Ref. 21. With the appearance and increase of losses, the field pattern changes qualitatively, the $C_6$ symmetry is broken. At an intermediate loss value ($\varepsilon'' = 0.20$), the symmetry of the field pattern becomes $C_2$ with a uniform field distribution both along the horizontal axis with two maxima and along the vertical axis with three maxima, Figure 6e,k. Finally, at losses corresponding to amorphous GST, the $C_2$ symmetry is retained, but the field along the horizontal and vertical axes ceases to be uniform. Moreover, it is clearly seen that the field distribution corresponding to quasi-BIC copies the field distribution in the low-frequency branch. This is not surprising when one look at Figure 3. In SCS spectra, the mode of the low-frequency branch is two orders of magnitude more intense than the quasi-BIC mode, its field in the cavity is decisive, and the quasi-BIC mode only slightly changes this distribution due to the interaction between them.

## 6. Two Mechanisms for Switching Optical Spectra

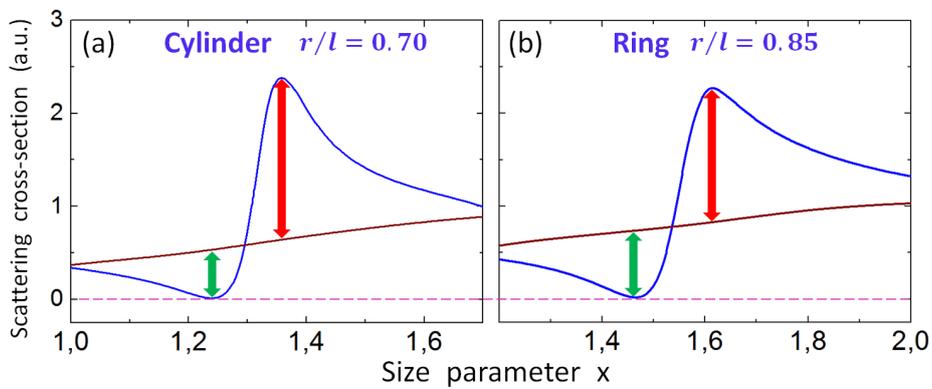

**Figure 7.** SCS spectra of the cylinder (a) and ring (b) in the amorphous (blue line) and crystalline (brown line) GST phases in the avoided crossing spectral region of the $TE_{0,2,0}$ and $TE_{0,1,2}$ modes. The lilac dotted line indicates zero scattering intensity. Normalized size parameter $x = r\omega/c$. TE-polarized incident wave.

Finally, we describe the effect of switching the SCS spectra upon the GST transition from the amorphous to the crystalline phase, taking into account real losses. Figure 7 shows the spectroscopic result of GST switching from the state with a complex permittivity 19.0+i0.42 to a state with a complex permittivity 41.6 + i17.5. Huge changes in the permittivity of the cylinder and ring from crystalline GST with respect to amorphous lead to a complete change in the SCS spectra, and as a result, in the discussed spectral region, we see only a monotonically varying background without any spectral features (brown line).

Thanks to **Figure 7**, we can describe two different mechanisms for switching the SCS spectra as a result of the amorphous – crystalline phase transition. The first mechanism, marked with red arrows in Figure 7, is associated with the disappearance of the resonant mode structure in the SCS spectra of the crystalline phase. More interesting is the second mechanism, marked in Figure 7 with green arrows. As noted above, the resonant mode structure of both the cylinder and the ring is described by Fano resonances and in the avoided crossing region by two Fano contours according to Formula (1). When material losses in the amorphous phase are taken into account, $\varepsilon'' = 0.42$, the band of the high-frequency quasi-BIC component is two orders of magnitude weaker than the low-frequency band, and to describe the SCS spectrum it is sufficient to use one Fano contour in Formula (1). A characteristic feature of the Fano contour is that its amplitude vanishes exactly at the frequency corresponding to the condition that the numerator of the Fano Formula vanishes $q_1+\Omega_1=0$. In the spectrum of amorphous GST, vanishing is observed at a normalized frequency of 1.22 for a cylinder and 1.43 for a ring. This second mechanism is more contrasting, since one of the states is close to zero.

## 7. Summary and Outlook

In this article, we present a general picture of the transformation of BIC in the Friedrich - Wintgen mechanism, depending on material losses. Carrying out calculations for two different resonators, the results of which fully correlate with each other, allow us to be confident in the validity of the conclusions. We succeeded in separation of the contributions of radiation and material losses in the scattering spectra, bringing the results to specific values. The value of the material quality factor for the amorphous GST was obtained, which is $Q_{mat} = 46$. In addition, we have determined the radiation quality factors for the cylinder ($Q_{rad}$=608) and ring ($Q_{rad}$=332) in the quasi-BIC regime of avoided crossing between the Mie-type TE$_{0,2,0}$ and Fabry-Perot-type TE$_{0,1,2}$ modes. Thus, the total quality factor is determined by material losses in the amorphous GST and does not differ significantly for the cylinder ($Q_{tot}^{\varepsilon''=0.42} = 43$) and the ring ($Q_{tot}^{\varepsilon''=0.42} = 41$).

As a result of calculations, we received a number of unexpected results. The biggest surprise turned out to be the behavior of two interacting modes, shown in Figures 2, 3. With an increase in losses far from the quasi-BIC region, both modes decrease in intensity in approximately the same way. However, for the aspect ratio corresponding to the minimum difference between the low frequency and high frequency branches, the quasi-BIC becomes practically dark mode while the low frequency band continues to be intense. In this case, the main effect is not associated with broadening, as one might expect, but with a decrease in the amplitude. It turned out to be possible to separate the line contours and isolate the quasi-BIC only as a result of the precision processing of the explosive spectrum according to the Fano Formula (1).

Our calculations have shown that material losses, which significantly reduce the *Q*-factor, in the investigated range from 0 to 0.42 have practically no effect on the mode coupling strength. Indeed, the calculated dependences of the low-frequency and high-frequency branches (green and red dependences shown in Figure 4a,f) practically do not change with the change in losses, that is, the Ruby splitting[48] remains unchanged. That is why in Figure 4 we present these dependences only for zero losses.

In addition, in the scattering spectra, we have demonstrated the effects of the phase transition from the amorphous GST phase to the crystalline phase and identified two mechanisms that can be used in practice. Particularly interesting is the mechanism due to the Fano resonance, when switching occurs between two signals, one of which is close to zero (green arrows in Figure 7).

In conclusion, our results demonstrate a simple and clear way to analyze losses in creating efficient resonant devices as a result of optimal combinations of three factors:

material, resonator geometry, and resonant photonic regime, one of which can be bound states in the continuum.

## Acknowledgements

This work was funded by the Russian Science Foundation (Project 20-12-00272).

## Conflict of Interest

The authors declare no conflict of interest

## Keywords

material losses, radiation losses, bound states in the continuum, $Ge_2Sb_2Te_5$

Received: ((will be filled in by the editorial staff))
Revised: ((will be filled in by the editorial staff))
Published online: ((will be filled in by the editorial staff))

[1] F. J. Garcia-Vidal, L. Martin-Moreno, T. W. Ebbesen, L. Kuipers, *Rev. Mod. Phys.* **2010**, *82*, 729.

[2] F. Koenderink, A. Alu, A. Polman, *Science* **2015**, *348*, 516-521.

[3] A.I. Kuznetsov, A. E. Miroshnichenko, Y. H. Fu, J. Zhang, B. Luk'yanchuk, *Sci. Rep.* **2012**, *2*, 492.

[4] J. C. Ginn, I. Brener, D. W. Peters, J. R. Wendt, J. O. Stevens, P. F. Hines, L. I. Basilio, L. K. Warne, J. F. Ihlefeld, P. G. Clem, M. B. Sinclair, *Phys. Rev Lett.* **2012**, *108*, 097402.

[5] A. B. Evlyukhin, A. B. Evlyukhin, S. M. Novikov, U. Zywietz, R. L. Eriksen, C. Reinhardt, S. I. Bozhevolnyi, B, N. Chichkov, *Nano Letters* **2012**, *12*, 3749.

[6] J. D. Jackson, *Classical Electrodynamics*, Wiley, New York, **1998**.

[7] A. Lagendijk, B. van Tiggelen, D. S.Wiersma, *Phys. Today* **2009**, *62*, 24.

[8] K. Inoue, K. Ohtaka, *Photonic Crystals: Physics, Fabrication and Applications*, Springer-Verlag, Berlin, **2004**.

[9] C. W. Hsu, B. Zhen, A. D. Stone, J. D. Joannopoulos, M. Soljačić, *Nat. Rev. Mater.* **2016**, *1*, 16048.

[10] J. von Neumann, E. Wigner, *Phys. Z.* **1929**, *30*, 465.

[11] C. W. Hsu, B. Zhen, J. Lee, S.-L. Chua, S. G. Johnson, J. D. Joannopoulos, M. Soljačić, *Nature* **2013**, *449*, 188–191.

[12] C. W. Hsu, B. Zhen, A. D. Stone, J. D. Joannopoulos, M. Soljačić, *Nature Reviews Mater.* **2016**, *1*, 16048.

[13] F. Monticone, A. Alu, *Phys. Rev. Lett.* **2014**, *112*, 213903.

[14] H. M. Doeleman, F. Monticone, W. den Hollander, A. Alù, A. F. Koenderink, *Nat. Photonics* **2018**, *12*, 397.

[15] F. Monticone, H. M. Doeleman, W. Den Hollander, A. F. Koenderink, A. Alù, *Laser Photonics Rev.* **2018**, *12* 1700220.

[16] S. I. Azzam, V. M. Shalaev, A. Boltasseva, A. V. Kildishev, *Phys. Rev. Lett.* **2018**, *121*, 253901.

[17] K. Koshelev, G. Favraud, A. Bogdanov, Y. Kivshar, A. Fratalocchi, *Nanophotonics*, **2019**, *8*, 725–745.

[18] S. I. Azzam, A. V. Kildishev, *Adv. Optical Mater.* **2021,** *9*, 2001469.

[19] S. Joseph, S. Sarkar, S. Khan, J. Joseph, *Adv. Optical Mater.* **2021**, *9*, 2001895.

[20] H. Friedrich, D Wintgen, *Phys. Rev. A* **1985**, *32*, 3231.

[21] M. V. Rybin, K. L. Koshelev, Z. F. Sadrieva, K. B. Samusev, A. A. Bogdanov, M. F. Limonov, Y. S. Kivshar, *Phys. Rev. Lett.* **2017**, *119*, 243901.

[22] A. A. Bogdanov, K. L. Koshelv, P. V. Kapitanova, M. V. Rybin, S. A. Gladyshev, Z. F. Sadrieva, K. B. Samusev, Y. S. Kivshar, M. F. Limonov, *Adv. Photonics* **2019**, *1* 016001.

[23] N. Solodovchenko, K. Samusev, D. Bochek, M. Limonov, *Nanophotonics,* **2021**, *10*, 0351.

[24] T. Cao, X. Zhang, W. Dong, L. Lu, X. Zhou, X. Zhuang, J. Deng, X. Cheng, G. Li, R. E. Simpson, *Adv. Opt. Mater.* **2018**, *6*, 1800169.

[25] W. Dong, Y. Qiu, X. Zhou, A. Banas, K. Banas, M. B. H. Breese, T. Cao, R. E. Simpson, *Adv. Opt. Mater.* **2018**, *6*, 1701346.

[26] Y. G. Chen, T. S. Kao, B. Ng, X. Li, X. G. Luo, B. Luk'yanchuk, S. A. Maier, M. H. Hong, *Opt. Express* **2013**, *21*, 13691.

[27] C. Choi, S.-Y. Lee, S.-E. Mun, G.-Y. Lee, J. Sung, H. Yun, J.-H. Yang, H.-O. Kim, C.-Y. Hwang, B. Lee, *Adv. Optical Mater.* **2019**, *7*, 1900171.

[28] A. V. Pogrebnyakov, J. A. Bossard, J. P. Turpin, J. D. Musgraves, H. J. Shin, C. Rivero-Baleine, N. Podraza, K. A. Richardson, D. H. Werner, T. S. Mayer, *Opt. Mater. Express* **2018**, *8,* 2264.

[29] A. B. Pevtsov, A. N. Poddubny, S. A. Yakovlev, D. A. Kurdyukov, V. G. Golubev, *J. Appl. Phys*. **2013**, *113,* 144311.

[30] S. A. Dyakov, N. A. Gippius, M. M. Voronov, S. A. Yakovlev, A. B. Pevtsov, I. A. Akimov, S. G. Tikhodeev, *Phys. Rev. B* **2017**, *96*, 045426.

[31] W. H. P. Pernice, H. Bhaskaran, *Appl. Phys. Lett*. **2012**, *101*, 171101.

[32] A.-K. U. Michel, P. Zalden, D. N. Chigrin, M. Wuttig, A. M. Lindenberg, T. Taubner. *ACS Photonics*, **2014**, *1*, 833–839.

[33] M. Odit, K. Koshelev, S. Gladyshev, K. Ladutenko, Y. Kivshar, A. Bogdanov, *Adv. Mater.* **2020**, *33*, 2003804.

[34] M. Minkov, I. A. D. Williamson, M. Xiao, S. Fan, *Phys. Rev. Lett.* **2018**, *121*, 263901.

[35] L. Vertchenko, C. DeVault, R. Malureanu, E. Mazur, A. Lavrinenko, *Laser Photonics Rev.* **2021**, *15*, 2000559.

[36] M. G. Silveirinha, *Phys. Rev. A* **2014**, *89*, 023813.

[37] K. Koshelev, G. Favraud, A. Bogdanov, Y. Kivshar, A. Fratalocchi, *Nanophotonics* **2019**, *8* 725–745.

[38] Z. Hayran, F. Monticone, *ACS Photonics* **2021**, *8*, 813-823.

[39] B. Li, J. Yao, H. Zhu, G. Ca, Q. H. Liu, *Optical Materials Express*, **2021**, *11*, 2359.

[40] W. Wang, H. Yan, L. Xiong, L. Zheng, *J. of Modern Optics*, **2021**, *68*, 699–706.

[41] Z. Han, F. Ding, Y. Cai, U. Levy, Nanophotonics **2021**, *10,* 1189–1196.

[42] D. V. Bochek, D. A. Yavsin, A. B. Pevtsov, K. B. Samusev, M. F. Limonov, *PNFA*, **2021**, *44*, 100906.

[43] W. Wełnic, A. Pamungkas, R. Detemple, C. Steimer, S. Blügel, M. Wuttig, *Nat. Mater.* **2006**, *5*, 56.

[44] K. Shportko, S. Kremers, M. Woda, D. Lencer, J. Robertson, M. Wuttig, *Nat. Mater.* **2008**, *7*, 653.

[45] P. Němec, A. Moreac, V. Nazabal, M. Pavlišta, J. Přikryl, M. Frumar, *J. Appl. Phys.* **2009**, *106*, 103509.

[46] K. Zhang, D. Li, *Electromagnetic Theory for Microwaves and Optoelectronics*, Springer, Berlin, **2008**.

[47] M. F. Limonov, *Adv. Opt. Photon*, **2021**, *13*, 703-771.

[48] M. F. Limonov, M. V. Rybin, A. N. Poddubny, Y. S. Kivshar. *Nature Photonics*, **2017**, *11*, 543-554.

[49] M. V. Rybin, K. B. Samusev, I. S. Sinev, G. Semouchkin, E. Semouchkina, Yu. S. Kivshar, M. F. Limonov, *Optics Express*, **2013**, *21*, 30107.

[50] M. V. Rybin, D. S. Filonov, P. A. Belov, Yu. S. Kivshar, M. F. Limonov, *Sci. Rep.* **2015**, *5*, 8774.

[51] E. E. Maslova, M. V. Rybin, A. A. Bogdanov, Z. F. Sadrieva, *Nanophotonics,* **2021**, *10*, 0475.